\documentclass[11pt,twoside]{article}
\usepackage[english,russian]{babel}
\usepackage{./asp2014}
\resetcounters

\begin{document}

\title{Structure of the shock at the surface of CTTS}
\author{Alexandr~Dodin
\affil{Sternberg Astronomical Institute, M.\,V. Lomonosov Moscow State University, Moscow, Russia; \email{dodin\_nv@mail.ru}}
}

\paperauthor{Alexandr Dodin}{dodin\_nv@mail.ru}{}{Sternberg Astronomical Institute, M.\,V. Lomonosov Moscow State University}{}{Moscow}{}{119234}{Russia}

\begin{abstract}
A steady-state structure and spectrum of the accretion zone at the surface of T Tauri stars are numerically simulated.
This modeling uses well-developed methods of the theory of stellar atmospheres with inclusion of the effects of the 
gas motion (advection and the doppler shift of the lines). The spectrum $I_{\nu}(\mu)$ and cooling rate 
of shocked gas are calculated. It is found that at the pre-shock densities $N_0>10^{13}$ cm$^{-3}$ helium and hydrogen lines 
show a red-shifted absorption, significantly broadened by the Stark effect.

\end{abstract}

\section{Introduction}
T Tauri stars are young stars which accrete matter from a protoplanetary disk. The accretion flow near the stellar surface is channeled by the magnetic field. The gas, moving with a velocity of about 300 km/s, is decelerated in the strong shock, converting all its kinetic energy into the heat. The near-surface accretion structure can be conventionally divided on three regions: I (pre-shock) -- free-falling gas, II (post-shock) -- cooling region -- hot gas slowly sink into the stellar atmosphere, III (hot spot) -- almost hydrostatic stellar atmosphere, heated by radiation from region I and II. One of the most realistic calculation of the region II was performed by \citet{Lamzin98}. However, these models assume a free escape of radiation from the post-shock that leads to an un-physical cooling down to zero. In this work we take into account self-absorption in the cooling region that allows us to overcome this problem and  make a smooth transition between II and III. Developed numerical methods allow us to investigate another problem: \citet{Dodin15} suggested that at high densities of infalling gas this gas can significantly absorb radiation from region III.
In this contribution we investigate spectrum of the hot spot, observed through the infalling gas, performing a self-consistent modeling I and III regions.

\section{Post-shock}
\subsection{Equations and methods}
A steady-state structure of the post-shock along the mass coordinate $m$ is described by equations: 
\begin{equation}\label{eq1}
    \rho V = u_0=const, \qquad P+\rho V^2 = const, \qquad u_0 \frac{dE}{dm}= HC,
\end{equation}
where
\begin{displaymath}
      E =  V^2/2 + P/\rho+\varepsilon_{ion}, \qquad 
      \varepsilon_{ion} = \frac{1}{\rho}\sum\limits_l {N_l E_l}, \qquad  
      HC = 4\pi \int{\chi_{\nu}(J_{\nu}-S_{\nu})d \nu}.
\end{displaymath}
Here $N_l, E_l$ are populations and energy for all levels of all ions, 
$\chi_{\nu}$, $J_{\nu}$, $S_{\nu}$ are the monochromatic absorption 
coefficient, mean intensity and source function, respectively.
$J_{\nu}$ is taken from the solution of radiative transfer equation, 
which is solved by the short characteristics method \citep{OK87} 
for the plane-parallel case.

A statistical equilibrium can be absent in the post-shock region \citep{Lamzin98}, 
therefore, to calculate the relative level populations $n_i$ of atomic levels, 
we should solve a corresponding system of differential-algebraic equations: 
\begin{equation}\label{eq2}
   u_0\frac{d n_i}{dm} = \sum\limits_{j=1}^{K} (R_{ji}+C_{ji})n_j - n_i \sum\limits_{j=1}^{K} (R_{ij}+C_{ij}), \quad i=(1,K-1), \quad
   \sum\limits_{i=1}^K n_i = 1,
\end{equation}
where $K$ is a number of the last considered level, namely, a ground state of the 
highest ionization stage of the element. $R_{ij}$, $C_{ij}$ are radiative and collisional rates.
The population of the last level $n_K$ 
can be expressed from the last equation through  $n_i,$ and then obtained stiff 
system of differential equations is solved by VODE code \citep*{Brown89}.
Finally, the level populations are calculated as $N_i = \xi_{el} N_A n_i,$
where $\xi_{el}$ is an abundance of element, $N_A$ is the total number 
density of atomic nuclei. Self-consistency of the radiation field and 
level populations is reached by the accelerated $\Lambda$-iterations.

To calculate temperature distribution in the post-shock region, which must satisfy the equations~(\ref{eq1}), the method, similar to well known $\Lambda$-correction, was applied 
\begin{equation}\label{eq3}
 C u_0 \frac{d \Delta T}{dm}=\frac{d HC}{d T} \Delta T + HC-u_0\frac{dE_0}{dm},
\end{equation}
where 
\begin{displaymath}
C=\frac{dE}{dT} = \left(1.5 + \frac{V_0-V}{V_0-2V} \right)\frac{k(N_A + N_e)}{\rho}+\frac{d \varepsilon_{ion}}{dT},
\end{displaymath}
$V_0$ denotes pre-shock velocity. This ODE was solved by Runge-Kutta method with boundary condition $\Delta T=0$ at $m=0.$

Generally speaking, the temperatures of electrons and ions behind the shock front are different. Suggested approach can be generalized for a two-temperature model, however there are no needs to do that, because we are interested in SED of the emerging radiation, while in the two-temperature region the gas emits negligible part of its thermal energy. Excluding this thin two-temperature structure from our model allows us to accelerate calculations and reduce the spatial grid.
Another similar problem is related with boundary conditions for the level populations (see Eq.~\ref{eq2}). It turns out that equilibrium populations immediately after the shock front are established very fast in the same two-temperature region, therefore, if we exclude this region, we should set equilibrium initial conditions for the level populations.

\subsection{Atomic data}
To simulate the structure and spectrum of the post-shock, we take into account the following atoms and ions:
H\,I-II, He\,I-III, C\,III-VII, N\,II-VIII, O\,II-VIII, Ne\,II-X, Mg\,IV-XII, Si\,IV-XII, S\,IV-XII, Fe\,V-XIV. The atomic model of each ion contains from 10 to 20 levels (superlevels). Radiative atomic data were taken from NORAD and TOPbase. Collisional coefficients are calculated using van-Regemorter and Seaton formulas.

\subsection{Results}
The structure of the post-shock for the parameters of the inflowing gas $V_0=300$ km\,s$^{-1}$ and $N_0= 10^{12}$ cm$^{-3}$ are depicted on the Figure~\ref{fig1}.
The ultraviolet part of SED and some profiles of arising emission lines are presented on the Figure~\ref{fig2}.
\articlefiguretwo{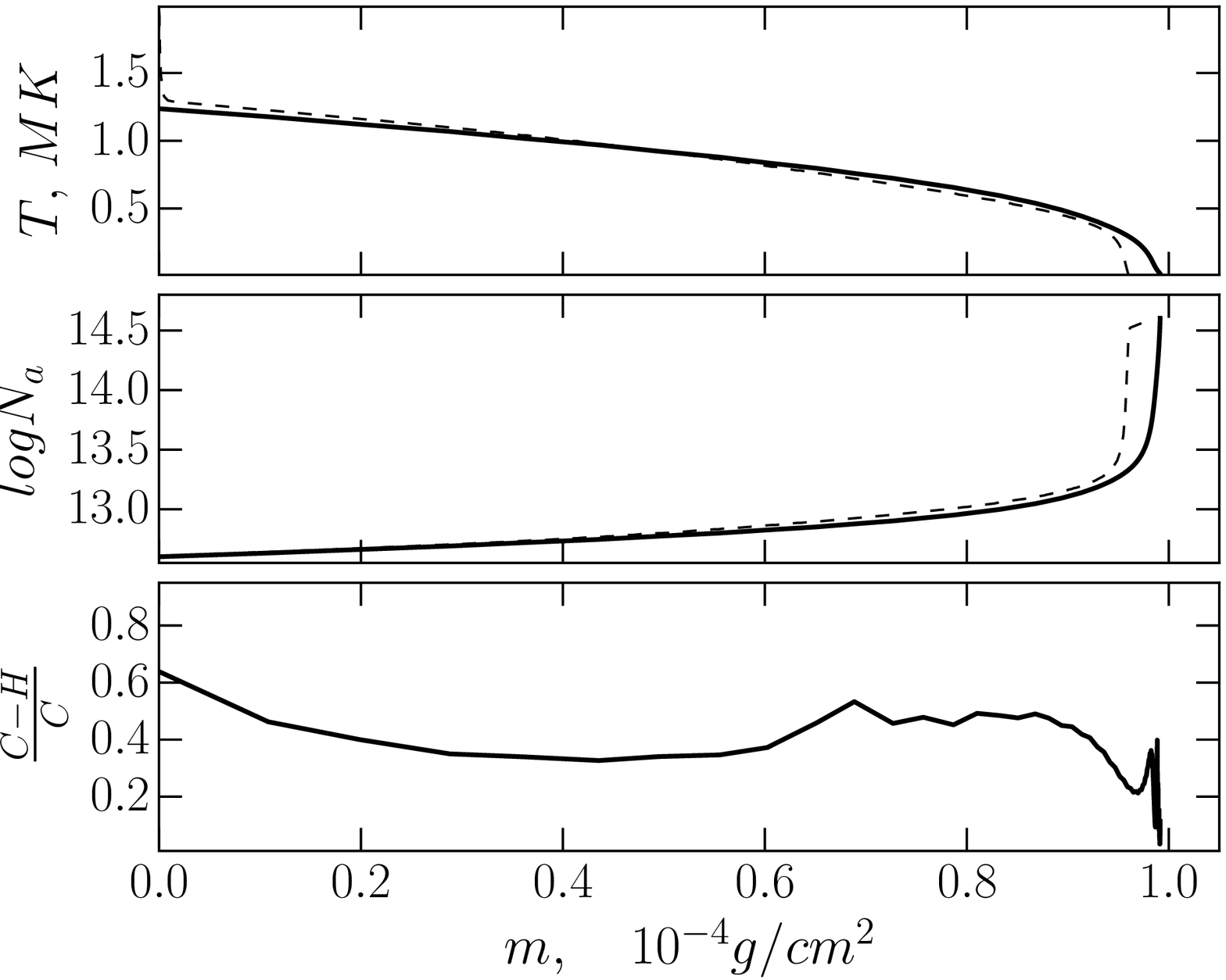}{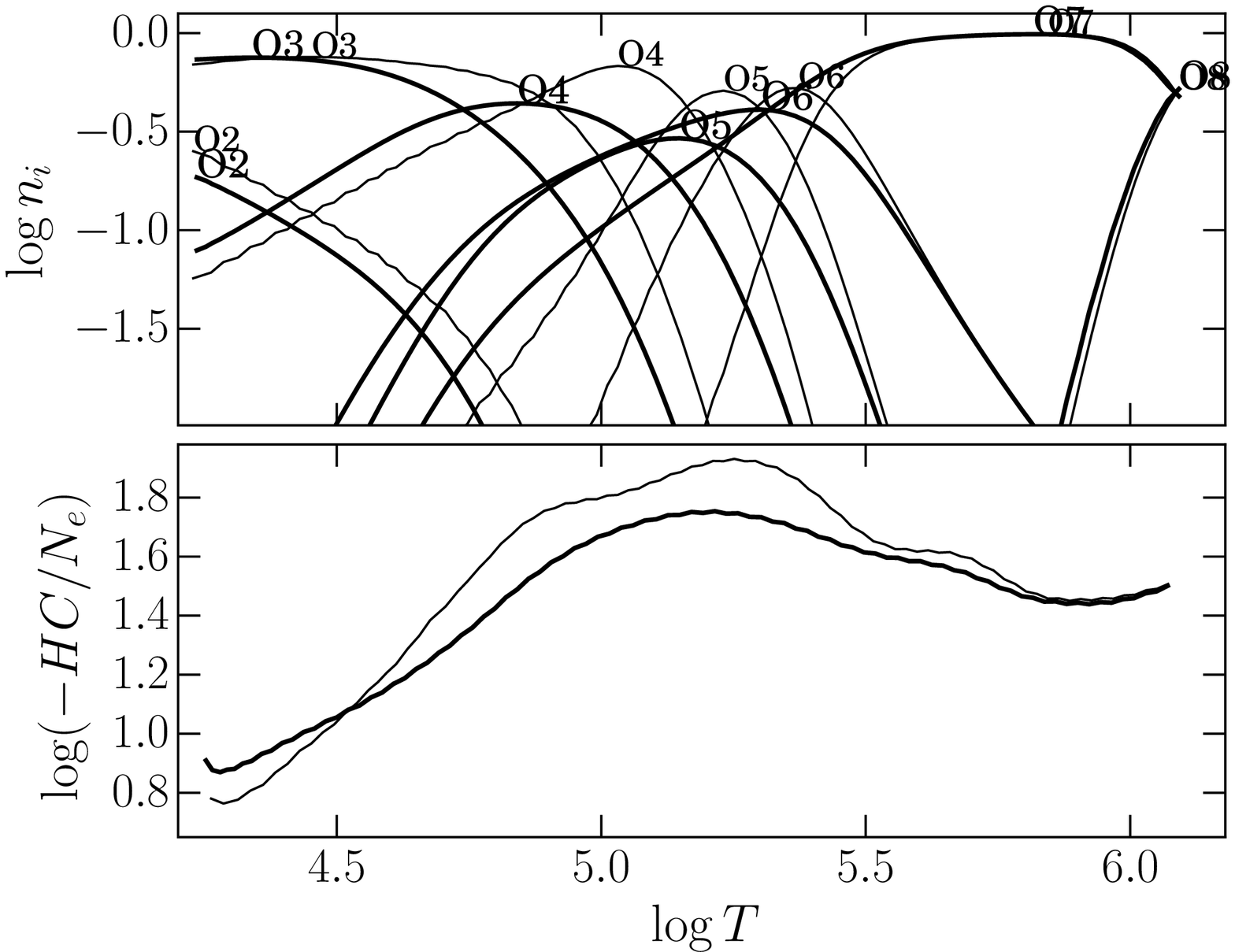}{fig1}{{\it Left:} the distribution  of temperature, number density and cooling (C) -- heating (H) disbalance are shown. Thick lines are for the present calculations, dashed lines are taken from \citet{Lamzin98}. {\it Right:} the distribution of ionization states of oxigen (upper panel) and "cooling function"{} (lower panel) are shown. Thin lines were obtained assuming a statistical equlibrium, while thick lines are calculated by solution of the Eq.~\ref{eq2}.}
\articlefiguretwo{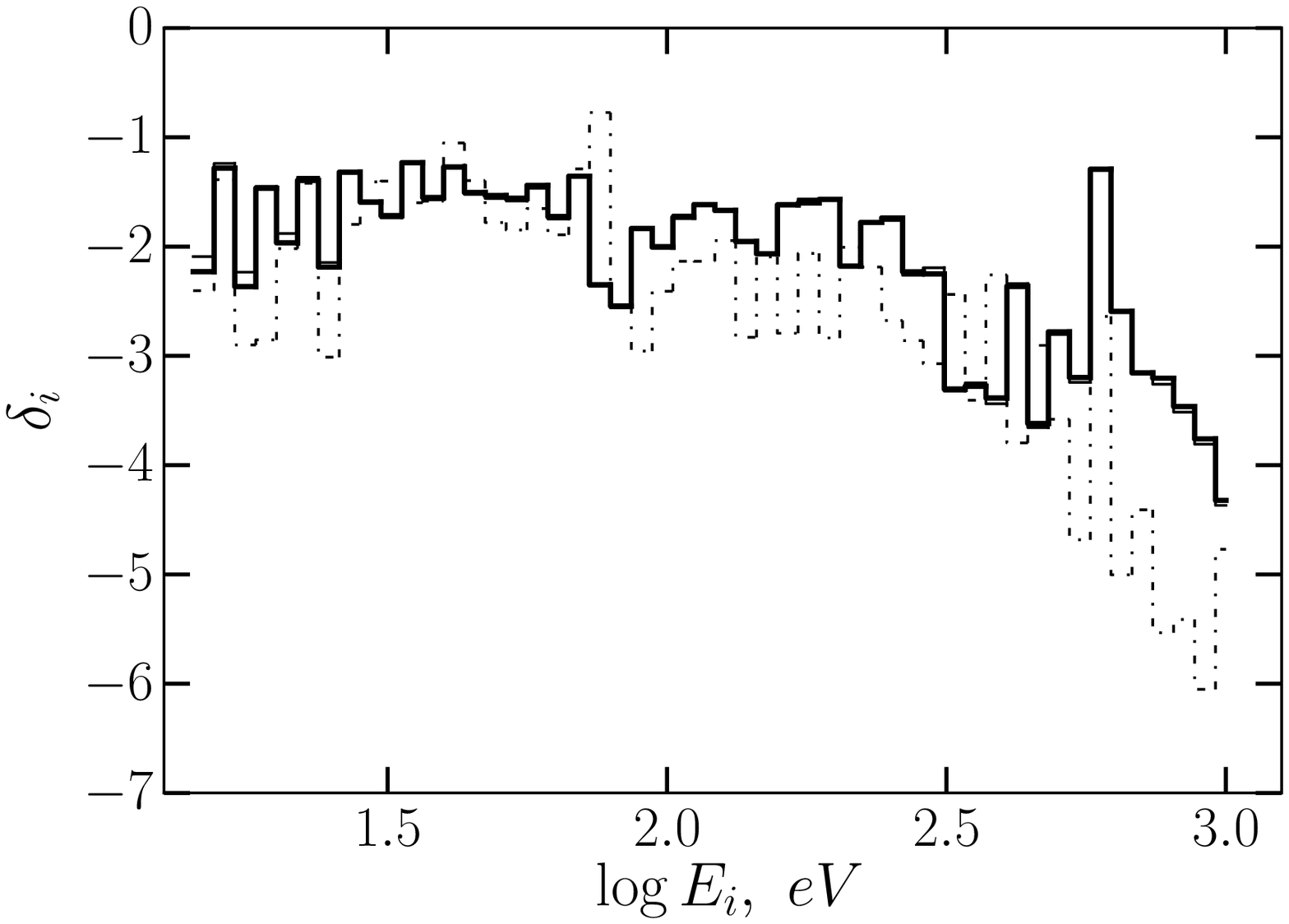}{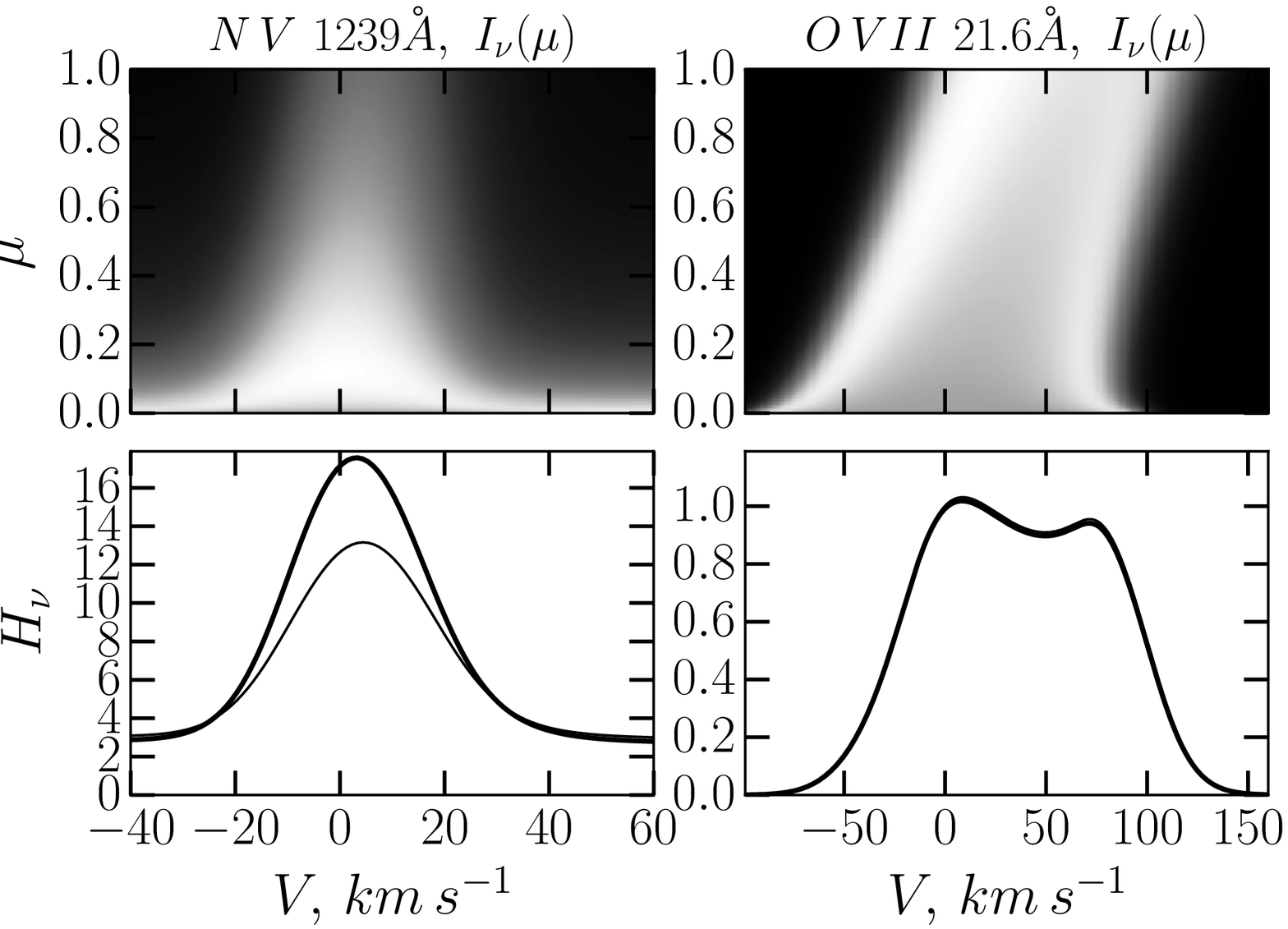}{fig2}{{\it Left:} SED is shown, $\delta_i=F_i/F_{acc},$ where $F_i$ is the flux in the energy interval between $E_i$ and $E_{i+1}$, $F_{acc}$ is the total accretion flux. Lines styles as for the Figure~\ref{fig2}. {\it Right:} the angular intensity distribution $I_{\nu}(\mu)$ for some lines (upper panels). Eddington flux $H_{\nu}$ for the same spectral lines in $10^{-6}$ erg s$^{-1}$ cm$^{-2}$ Hz$^{-1}$ sr$^{-1}$ (lower panels).}

\section{Hot spot and pre-shock}
Atomic data and methods for simulation of the structure of the hot spot are described in \citet{Dodin15}. The main improvement of the current work is self-consistent calculation of the infalling gas, using the same approach as in the case of the hot spot. The infalling gas is considered as a plane-parallel slab with a constant velocity $V_0$ and a number density $N_0.$ 
The structure of the hot spot and infalling gas is calculated simultaneously taking into account effects, caused by velocity, i.e. doppler shifts and advection.
The advection is calculated in the same manner as in the post-shock (see Eq.~\ref{eq2}, \ref{eq3}). However, the simulations show that the doppler shifts and advection do not change results significantly, and they can be neglected in qualitative consideration. Results of calculations for some spectral lines are shown on the Figure~\ref{fig3}.
It should be noted that at high densities $N_0$ the radiation from the hot spot is strongly absorbed by the infalling gas, in which red-shifted absorption lines are formed. Both central and red-shifted components are significantly broadened  by the Stark effect.
\articlefigure{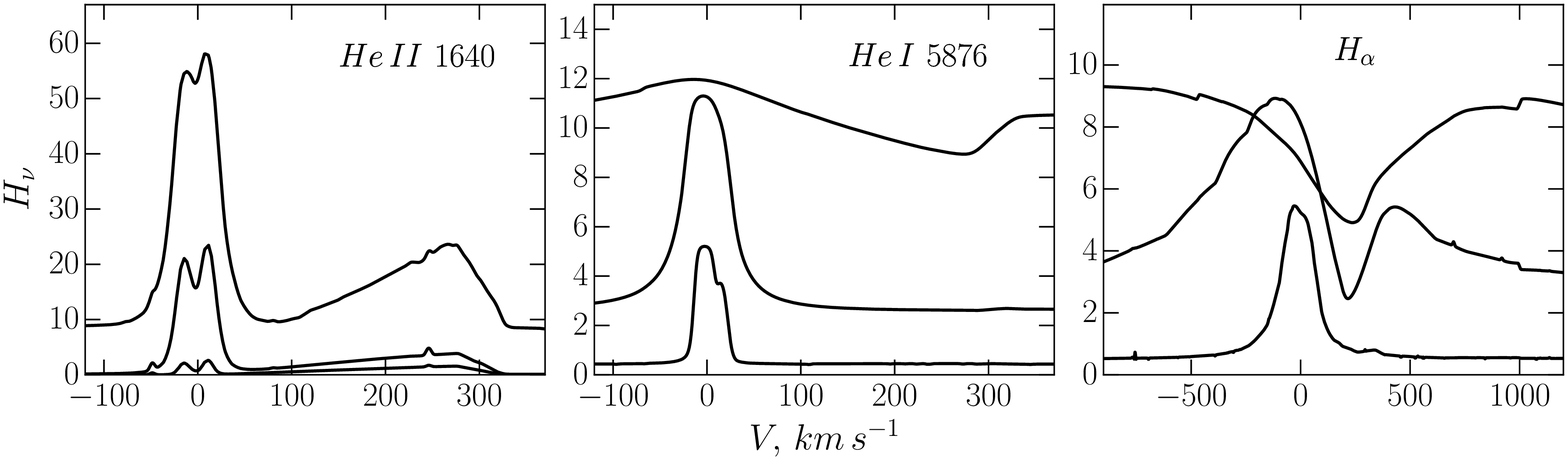}{fig3}{Profiles of some lines, arising in the hot spot and pre-shock, at various $\log N_0= 12,13,14$. Other parameters are $V_0=300$ km s$^{-1}$, $T_{eff}=4500$~K, $\log g = 4.0.$ $H_{\nu}$ -- Eggington flux in $10^{-5}$ erg s$^{-1}$ cm$^{-2}$ Hz$^{-1}$ sr$^{-1}.$ }

\acknowledgements This study was partially supported by RFBR, research project No. 16-32-00794 mol\_a.

\bibliography{aspdodin}

\begin{thebibliography}{}
\expandafter\ifx\csname natexlab\endcsname\relax\def\natexlab#1{#1}\fi
\expandafter\ifx\csname url\endcsname\relax
  \def\url#1{\texttt{#1}}\fi
\expandafter\ifx\csname urlprefix\endcsname\relax\def\urlprefix{URL }\fi
\providecommand{\eprint}[2][]{\url{#2}}

\bibitem[{Brown et~al.(1989)Brown, Byrne, \& Hindmarsh}]{Brown89}
Brown, P.~N., Byrne, G.~D., \& Hindmarsh, A.~C. 1989, SIAM J. Sci. Stat.
  Comput., 10, 1038

\bibitem[{Dodin(2015)}]{Dodin15}
Dodin, A.~V. 2015, Astronomy Lett., 41, 196

\bibitem[{Lamzin(1998)}]{Lamzin98}
Lamzin, S.~A. 1998, Astronomy Rep., 42, 322

\bibitem[{Olson \& Kunasz(1987)}]{OK87}
Olson, G.~L., \& Kunasz, P.~B. 1987, Journ. Quant. Spectrosc. Radiat.
  Transfer., 38, 325

\end{thebibliography}
\end{document}